\documentclass[showpacs,aps,pre,preprint]{revtex4}
\usepackage{graphicx}
\usepackage{amsfonts}

\def\eps{\varepsilon}

\begin{document}
\title{Dynamical localization of matter wave solitons in managed barrier potentials}
\author{Fatkhulla Kh. Abdullaev\dag
 and  Josselin Garnier\ddag
 }
\affiliation{\dag\ Physical-Technical Institute of the Uzbekistan Academy of
Sciences, 700084, Tashkent-84, G.Mavlyanov str.,2-b, Uzbekistan and
Instituto de Fisica Teorica, UNESP, Rua Pamplona, 145, Sao Paulo, Brasil\\
\ddag\ Laboratoire de Probabilit\'es et Mod\`eles
Al\'eatoires \& Laboratoire Jacques-Louis Lions, Universit{\'e}
Paris VII, 2 Place Jussieu, 75251 Paris Cedex 5, France}

\begin{abstract}
The bright matter wave soliton propagation through a barrier with
a rapidly oscillating position is investigated. The averaged over
rapid oscillations Gross-Pitaevskii (GP) equation is derived. It
is shown that the soliton is dynamically trapped by the effective double-barrier.
 The analytical predictions for the soliton effective dynamics
is confirmed by the numerical simulations of the full GP equation.
\end{abstract}

\maketitle

\section{Introduction}
The tunneling  of quantum particles is one of the fundamental
problems of quantum mechanics \cite{Landauer}. In connection with
this problem, the propagation of matter wave packets through different types
of potentials has recently attracted a lot of attention.
In particular it is interesting to study the transmission through
time-dependent barriers. Indeed, experiments with
dynamical tunneling of cold atoms in time-dependent barriers
of the multiplicative form $f(t)V(x)$
show the possibility for the tunneling period
control \cite{Hensinger,Steck}. The theory is developed for
linear wavepackets in Ref.~\cite{Averbukh}.
Soliton propagation through such a barrier is investigated in
Ref.~\cite{AG}. Another interesting problem is to consider a barrier with
a rapidly oscillating position, i.e. a potential of the form
$V(x-f(t))$. In the  linear regime of propagation
resonant effects are exhibited in the transmission process
through an  oscillating barrier \cite{Chiofalo,Embriaco}. It is therefore
relevant to investigate the propagation of {\it a
nonlinear wave packet} through an oscillating barrier. We
consider this problem for a bright matter wave soliton
in an attractive Bose-Einstein condensate (BEC) in the presence
of a barrier potential with an oscillating position.
Such a barrier can be achieved
by using a laser beam with a blue-detuned far-off-resonant
frequency. This generates a repulsive potential
and the laser sheet plays the role of a mirror for cold atoms.
Such a laser sheet has been used
to study the bouncing of a BEC cloud off a mirror \cite{Bongs}.
By moving the position of the mirror we obtain an oscillating barrier.

In this work we will study the dynamical trapping of soliton in
BEC with a rapidly oscillating barrier potential.  The averaged
over rapid oscillations  Gross-Pitaevskii  (GP) equation will be
derived. Applying a variational approach to this equation, we will
analyze the conditions for soliton trapping by the effective
potential. The analytical predictions are checked by numerical
simulations of the full GP equation.

\section{The averaged Gross-Pitaevskii equation}
The BEC wavefunction $\psi$ in a quasi one-dimensional geometry is
described by the GP equation
$$
i\hbar\psi_{T} + \frac{\hbar^2}{2m}\psi_{XX} - V_{p} (X-f_p(T) ,
T) \psi    - g_{1D}|\psi|^2 \psi=0,
$$
where $V_p(X,T)$ is a barrier potential.
Here $g_{1D} = 2\hbar a_s
\omega_{\perp}$ is the one-dimensional mean field nonlinearity
constant,  $a_s$ is the atomic scattering length,
$\omega_{\perp}$ is the transverse oscillator period,
and $a_{\perp} = \sqrt{\hbar/(m\omega_{\perp})}$ is the transverse
oscillator length.
Below we  consider the case of BEC with attractive interactions,
i.e. $a_s < 0$.
To avoid collapse in the attractive BEC, the condition $|a_s|N/a_{\perp} <
0.67$ should be satisfied,
with $N$ the number of atoms.
The barrier potential has an oscillating position described by
the zero-mean time-periodic function $f_p(T)$ with period $T_p$.
We consider the case of a fast-moving potential
in the sense that  $\eps := \omega_\perp T_p \ll 1$.
We also allow the barrier potential profile $V_p(X,T)$ to vary slowly in time,
i.e. with a  rate  smaller than $\omega_\perp$.
Introducing the standard dimensionless
variables
$$
x = \frac{\sqrt{2}X}{a_{\perp}}, \  \  \ t = \omega_{\perp}T,\ \ \  u =
\sqrt{|a_s|}\psi, \ \ \  V = \frac{V_p}{\hbar\omega_{\perp}},
$$
we obtain the
GP equation with fast moving potential
\begin{equation}
\label{eq:nls0} i u_t +u_{xx} - V \left(x-f(\frac{t}{\eps}) , t
\right) u  + 2 |u|^2 u =0,
\end{equation}
where
$$
f(t) = \frac{\sqrt{2}}{a_\perp}  f_p( T_p t)
$$
is a
periodic function with period $1$ and mean $0$.
The small parameter $\eps$ describes the
fast oscillation period in the dimensionless variables.
We look for the solution in the form
\begin{equation}
\label{eq:expand0}
u(x,t) = u^{(0)}(x,t,\frac{t}{\eps}) + \eps u^{(1)}(x,t,\frac{t}{\eps})  + \eps^2 u^{(2)}(x,t,\frac{t}{\eps}) + \cdots
\end{equation}
where $u^{(0)}$, $u^{(1)}$, $\ldots$ are  periodic in the argument $\tau=t/\eps$.
We substitute this ansatz into Eq.~(\ref{eq:nls0})
and collect the terms with the same powers
of $\eps$.
We obtain the hierarchy of equations:
\begin{eqnarray}
\label{hier0}
&& i u^{(0)}_\tau = 0, \\
\label{hier1}
&& i u^{(0)}_t +u^{(0)}_{xx}- V\left(x-f(\tau) ,t \right) u^{(0)}
+  2 |u^{(0)}|^2 u^{(0)} = -i u^{(1)}_\tau, \\
\label{hier2} && i u^{(1)}_t + u^{(1)}_{xx} - V\left(x-f(\tau) ,t
\right) u^{(1)} +  4 \bigl( u^{(0)} \bigr)^2\overline{u^{(1)}} 
+  2 \bigl| u^{(0} \bigr|^2
u^{(1)} =
 - i u^{(2)}_\tau.
\end{eqnarray}
The first equation (\ref{hier0})
shows that the leading-order term $u^{(0)}$
does not depend on $\tau$.
The second equation (\ref{hier1})
gives the compatibility equation for the existence
of the expansion (\ref{eq:expand0}):
$$
 i u^{(0)}_t +u^{(0)}_{xx}- \left< V\left(x-f(\cdot) ,t \right)\right> u^{(0)}  + 2 |u^{(0)}|^2 u^{(0)} =
 0,
$$
where $\left< \cdot \right>$ stands for an average in $\tau$.
Denoting
\begin{equation}
V_{\rm eff}(x) = \int_0^1 V\left(x-f(\tau) ,t \right) d\tau
\end{equation}
we obtain the averaged GP equation
\begin{equation}
\label{eq:avenls} iu_t +u_{xx}- V_{\rm eff} (x,t)u  + 2 |u|^2 u
=0.
\end{equation}
The equation (\ref{hier2}) allows us to compute the first order correction,
whose amplitude is of order $\eps$.

If, for example, $f(\tau)= a \sin(2 \pi \tau)$,
then a straightforward calculation shows that
\begin{equation}
V_{\rm eff}(x,t) = K_{\rm eff} * V(x,t),
\end{equation}
where $*$ stands for the convolution product (in $x$) and $K_{\rm eff}$ is given by
\begin{equation}
K_{\rm eff}(x)= \frac{1}{\pi} \frac{1}{\sqrt{a^2-x^2} }{\bf
1}_{(-a,a)}(x). \label{eq:expresK}
 \end{equation}
 If, moreover, $V(x,t)$ is a delta-like potential centered at a position $x_0(t)$
 that moves slowly in time,
 then
 $$
 V_{\rm eff}(x,t) = K_{\rm eff}(x-x_0(t))
 $$
 is a moving double-barrier potential.
By this way, one can generate a moving trap potential to
manage the soliton position.

\section{Variational approach for the soliton motion}
In absence of effective potential $V_{\rm eff}=0$ the NLS equation
(\ref{eq:avenls}) supports soliton solutions of the form
$$
u_0(x,t) = 2\nu \frac{e^{i 2\mu (x-x_s(t) ) + 4i (\mu^2+\nu^2)t }
}{\cosh[2 \nu(x-x_s(t)) ]}
$$
with $x_s(t)=x_s(0)+4 \mu t$.
In the presence of a stationary potential $V_{\rm eff}(x)$,
the soliton dynamics can be studied
by perturbation techniques.
Applying the first-order perturbed Inverse Scattering Transform
theory \cite{Karpman}, we obtain the
system of equations for the soliton amplitude and velocity
$$
\frac{d \nu}{dt}=0, \ \ \ \ \ \ \frac{d\mu}{dt} = -\frac{1}{4 \nu}
W_\nu'(x_s), \ \ \ \ \ \ \frac{d x_s}{dt} = 4 \mu ,
$$
where the prime stands for a derivative with respect to $x$ and
$W_\nu$  has the form
\begin{eqnarray}
\nonumber
W_\nu (x) &=&  \nu \int_{-\infty}^{\infty} \frac{1}{\cosh^2(z)} V_{\rm eff}
\bigl(x-\frac{z}{2\nu}
\bigr) dz\\
\label{def:wl}
&=& K_\nu* V_{\rm eff}(x)\, ,\\
K_\nu(x) &=& \frac{2\nu^2}{\cosh^2(2\nu x)}\, .
 \end{eqnarray}
Therefore, we can write the effective equation describing the dynamics of the soliton
center as a quasi-particle moving in the effective potential $W_{\nu_0}$:
\begin{equation}
\label{eq:quasiparteq1}
\nu_0 \frac{d^2 x_s}{dt^2} = - W_{\nu_0}'( x_s),
\end{equation}
where $4\nu_0$ is the mass (number of atoms) of the initial
soliton. This system has the integral of motion
\begin{equation}
\label{ener0} \frac{\nu_0}{2} \left(\frac{dx_s}{dt}\right)^2 +
W_{\nu_0}(   x_s) = 8 \nu_0 \mu_0^2,
\end{equation}
where $4\mu_0$ is the velocity of the initial soliton. Note that
this approach gives the same result as the adiabatic perturbation
theory for solitons that is a first-order method as well. This
adiabatic perturbation theory was originally introduced for
optical solitons \cite{kivshar89,Abd93} and it was recently
applied to matter wave solitons
 \cite{kevrekidis}.

\begin{figure}
\begin{center}
\begin{tabular}{c}
\includegraphics[width=6.4cm]{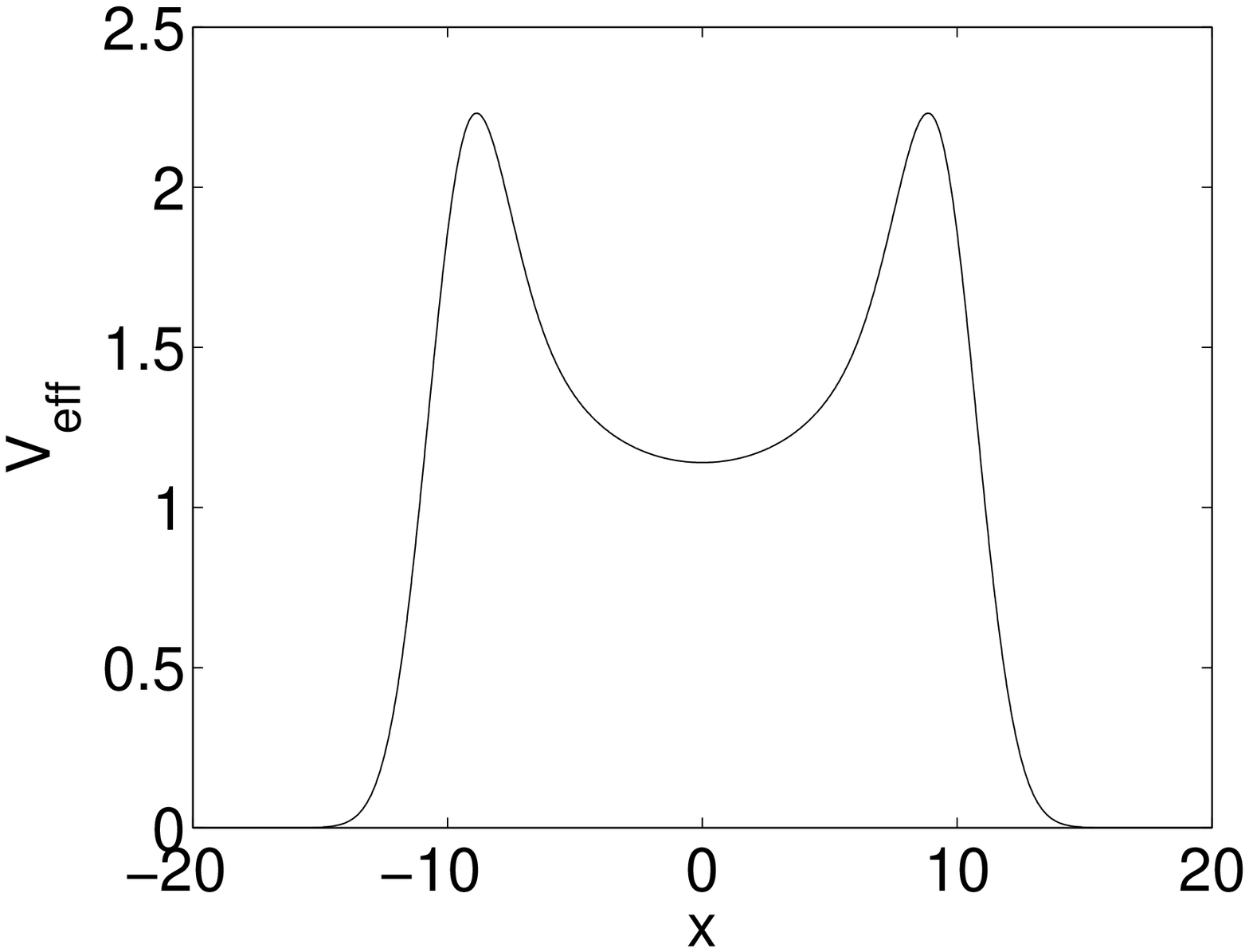}
\includegraphics[width=6.4cm]{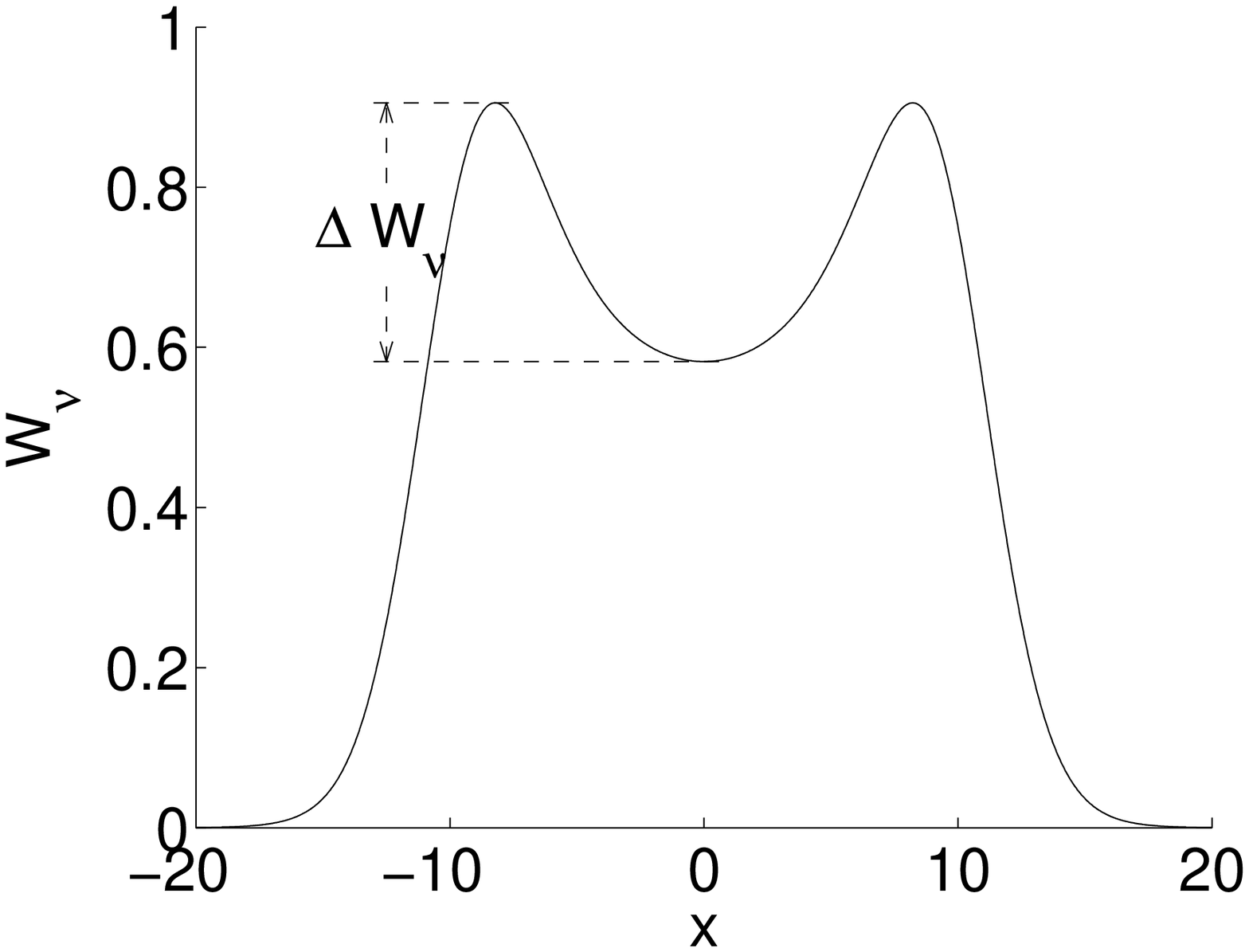}
\end{tabular}
\vspace*{-0.25in}
\end{center}
\caption{\label{fig0} Effective potentials $V_{\rm eff}(x)$
for the averaged equation (\ref{eq:avenls}) and $W_\nu(x)$
for the quasi-particle motion (\ref{eq:quasiparteq1}). Here $\nu=0.25$,
$V(x)=10 \exp(-x^2/4)$, and $f(\tau)=10 \sin(2 \pi \tau)$.}
 \end{figure}

The quasi-particle effective potential $W_\nu$ is given by
$$
W_\nu(x) = (K_\nu * K_{\rm eff}) *V(x).
$$
In the case where $f(\tau) = a\sin(2 \pi \tau)$,
the kernel $K_{\rm eff}$ is given by (\ref{eq:expresK}) and the Fourier transform
of the quasi-particle potential
has the  explicit form
$$
\hat{W}_\nu(k) = \frac{\pi k J_0(ka)}{2 \sinh(\frac{\pi k}{4 \nu} )} \hat{V}(k) \, ,
$$
where $\hat{V}$ is the Fourier transform of the reference potential $V$.
The effective quasi-particle potential is plotted in Fig.~\ref{fig0}
for a Gaussian barrier potential $V$.
The potential $W_\nu$ has a local minimum at $x=0$ between  two
global maxima that are close to $x = \pm a$.
The trap   amplitude is
$$
\Delta W_\nu = \max_{x} W_{\nu}(x)  -W_{\nu}(0).
$$
Let us consider an input soliton at $x=0$ with parameters $(\nu_0,\mu_0)$.
If the initial soliton velocity is large enough,
then the soliton escapes the trap. There is a critical value $\mu_c$ for
the initial soliton velocity parameter $\mu_0$ defined by
$$
\mu_c^2 = \frac{\Delta W_{\nu_0}}{8 \nu_0}
$$
that determines the type of motion: \\
- If $|\mu_0|<\mu_c$ then the soliton is trapped.
Its motion is oscillatory between the 
positions
$\pm x_f$ defined by
\begin{equation}
W_{\nu_0}(x_f) - W_{\nu_0}(0) = 8 \nu_0^2 \mu_0^2
\end{equation}
- If $|\mu_0| > \mu_c$, then the soliton motion is unbounded.
It escapes the trap and reaches the asymptotic velocity
parameter $\mu_a$ given by
\begin{equation}
\label{eq:asyvel}
\mu_a^2 = \frac{W_{\nu_0}(0)}{8 \nu_0} +\mu_0^2
\end{equation}
which shows that the transmitted soliton velocity is larger than
the initial soliton velocity.

As we shall see in the numerical simulations,
if the initial soliton parameters are close to the critical case
$|\mu_0| \sim \mu_c$, then
radiation effects become non-negligible.
The construction of an efficient trap requires
to generate a barrier potential $V$ that is high enough so that
$\Delta W_{\nu_0}$ is significantly larger than $8 \nu_0^2 \mu_0$.

Finally, if the potential $V$ is not stationary but
has an explicit time-dependence, such as a drift,
then the adiabatic equations derived in this section still hold true
if the time-dependence is slow enough.

\section{Numerical simulations and experimental predictions}

In this section we show the motion of a soliton with initial
parameters $\nu_0=\mu_0=0.25$ (the soliton velocity is $1$).
The reference barrier potential is $V(x) =  10 \exp(-x^2/4)$.
Note that in this case, $\Delta W_{\nu_0} \simeq 0.323$
and $\mu_c \simeq 0.4$.
Therefore, the quasi-particle approach
predicts trapping.

\begin{figure}
\begin{center}
\begin{tabular}{c}
\includegraphics[width=6.4cm]{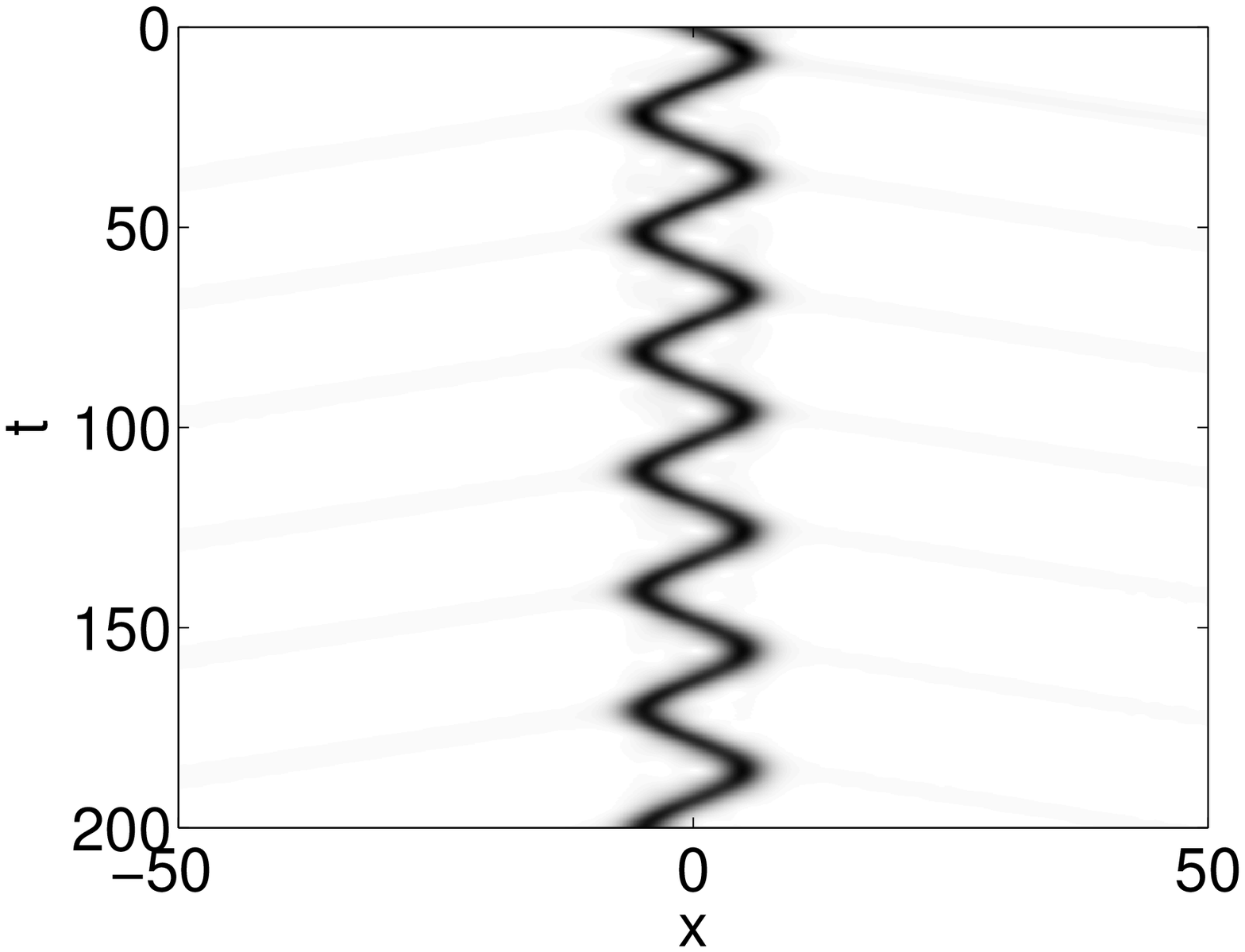}
\includegraphics[width=6.4cm]{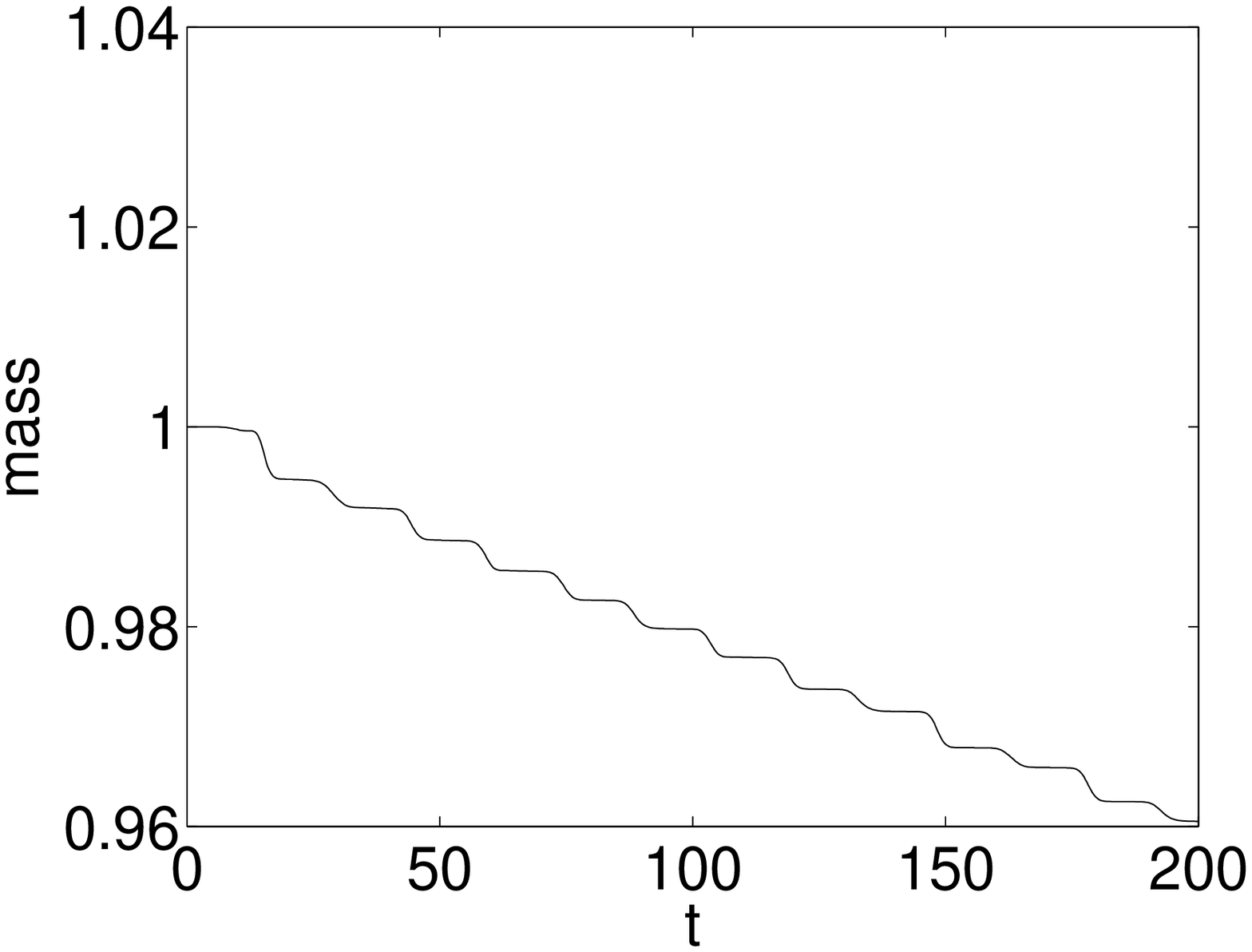} \\
\includegraphics[width=6.4cm]{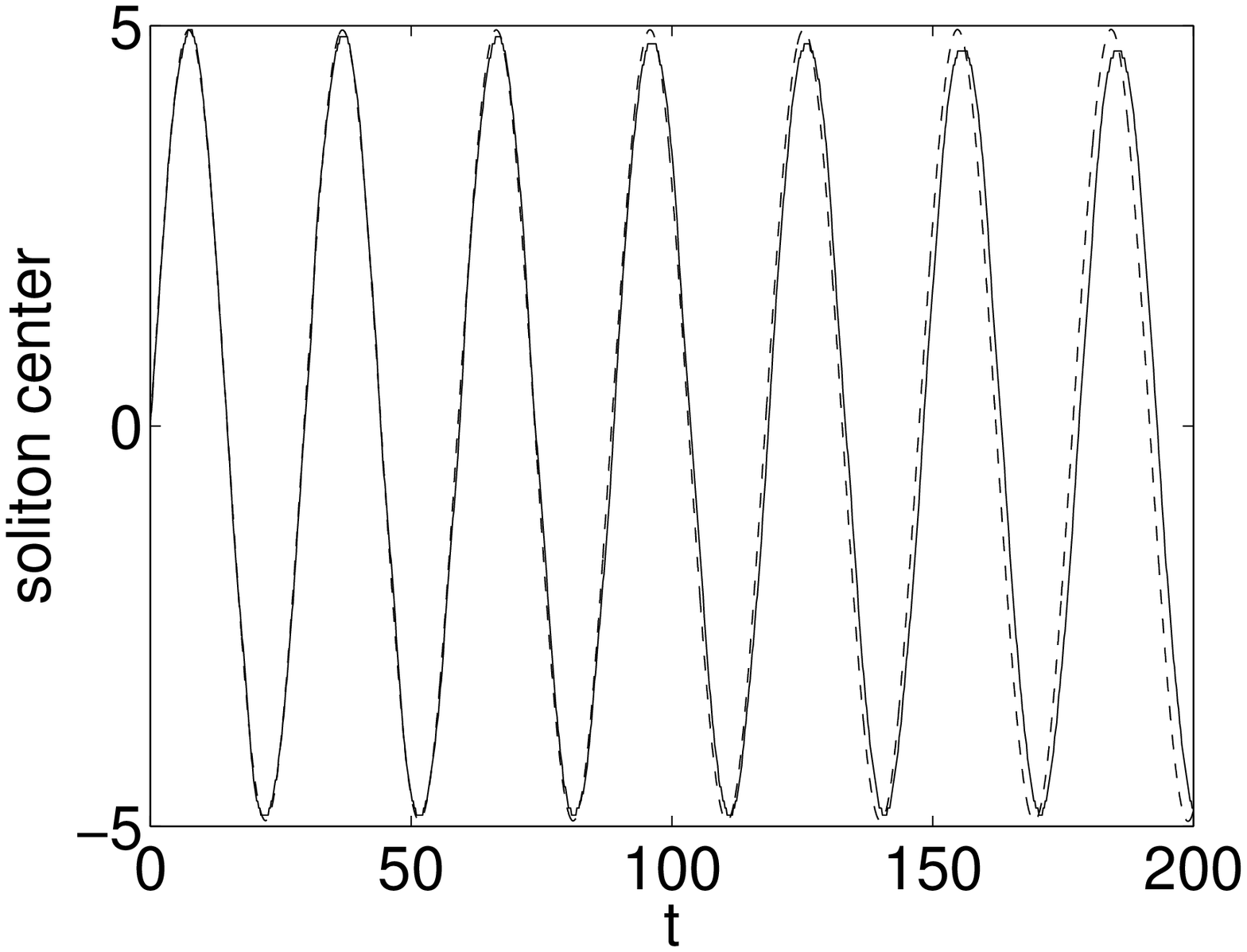}
\end{tabular}
\vspace*{-0.25in}
\end{center}
\caption{\label{fig2} Soliton profile in a  narrow trap $V[x - 10\sin(50 t)]$
 (left picture).
 Mass decay (right picture).
 Bottom picture:
 Soliton motion predicted by the quasi-particle approach (dashed)
 and  observed in the numerical simulations (solid).}
\end{figure}

In Fig.~\ref{fig2} the modulated potential is
$V[x-10\sin(50 t) ]$ (stationary trap).
Here $\eps = 2 \pi /50 \simeq 0.126$
and we check by full numerical simulations of the GP equation (\ref{eq:nls0})
that $\eps$ is small enough to ensure the validity of the theoretical
predictions based on the averaged GP equation (\ref{eq:avenls}).
We plot the soliton profile and the soliton mass, which shows that
radiative effects are very small.
We also compare the soliton motion obtained by the  numerical simulations
with the one predicted by the quasi-particle approach,
which shows very good agreement.

\begin{figure}
\begin{center}
\begin{tabular}{c}
\includegraphics[width=6.4cm]{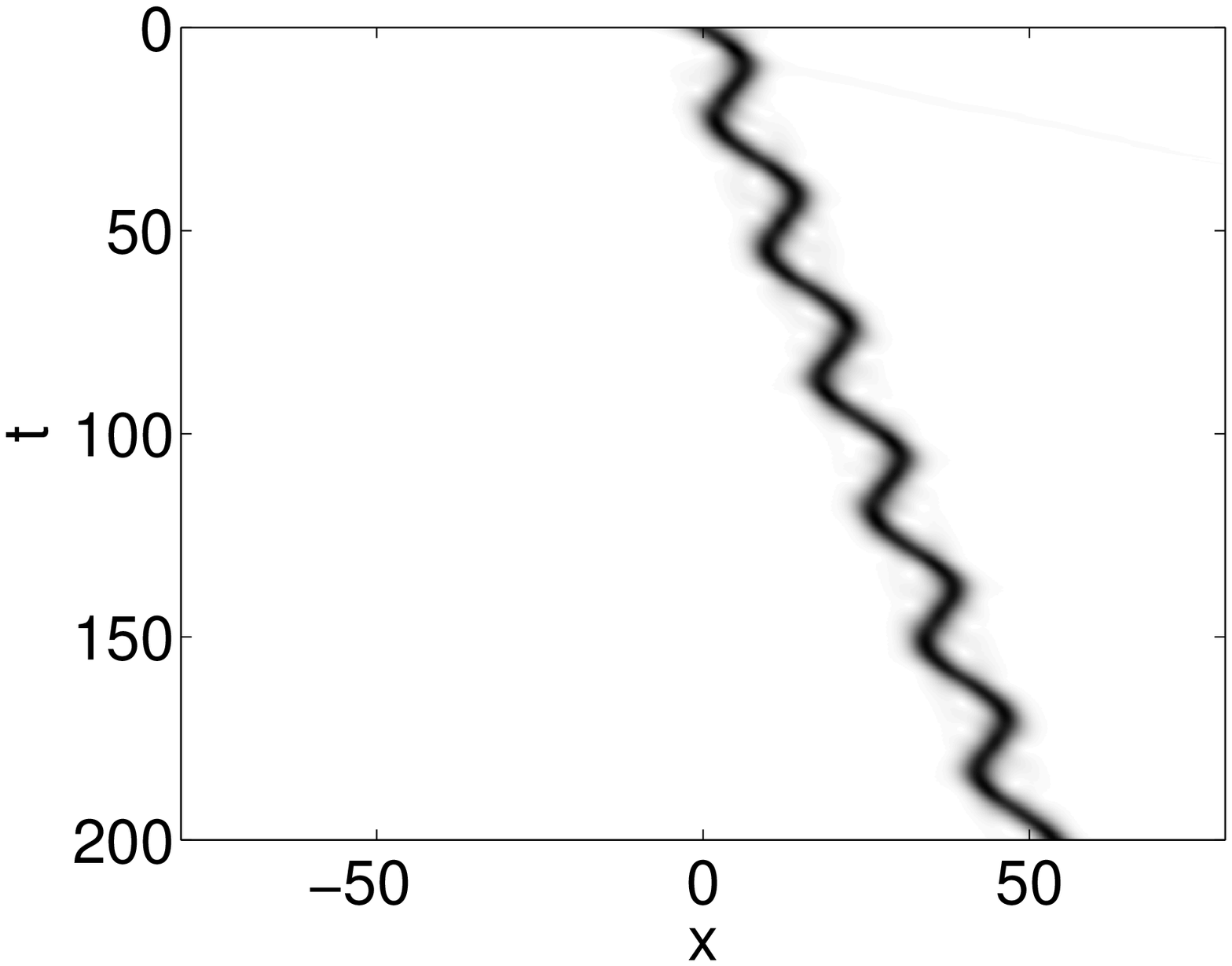}
\includegraphics[width=6.4cm]{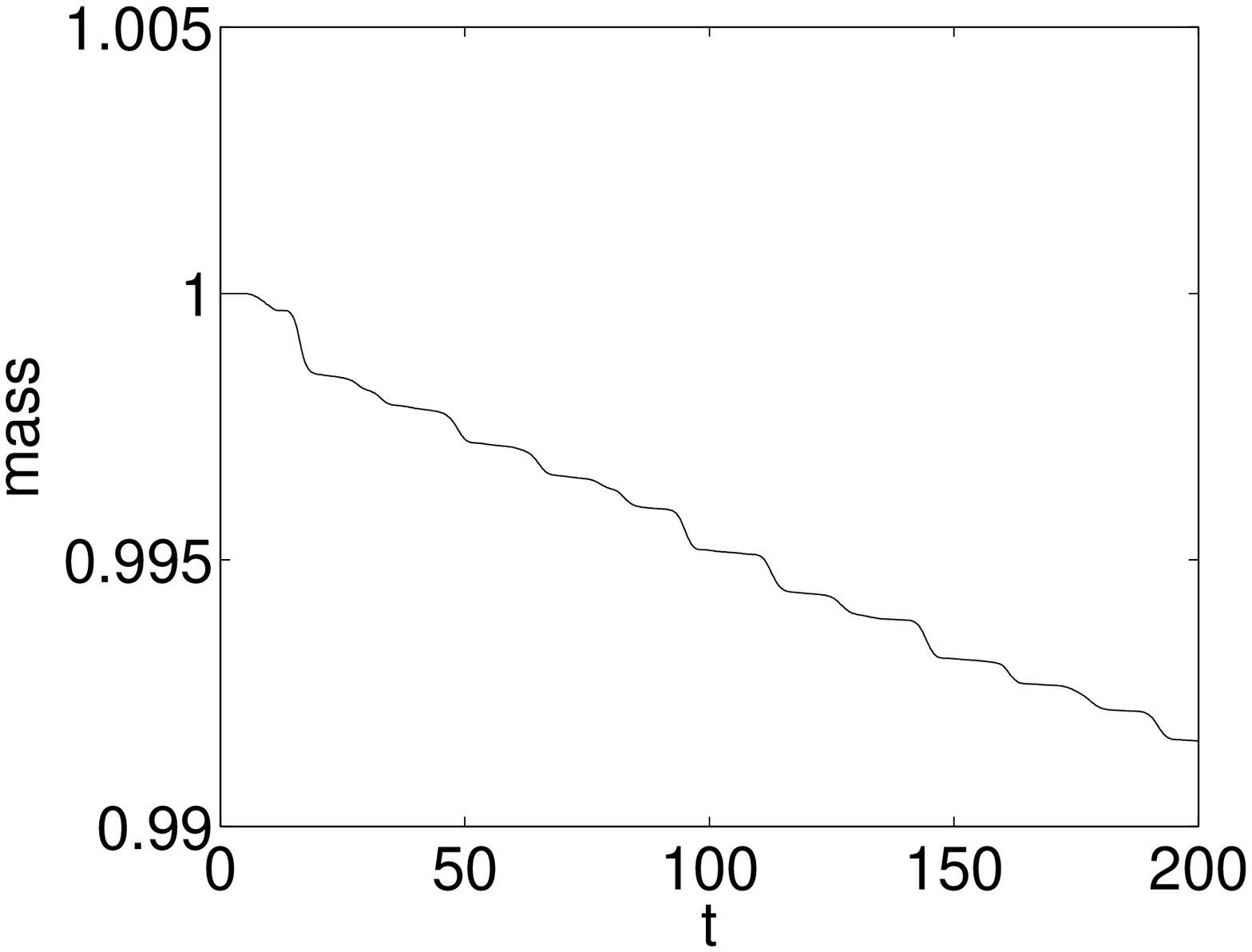}
\end{tabular}
\vspace*{-0.25in}
\end{center}
\caption{\label{fig3} Soliton profile in a narrow moving  trap $
 V[x - 10\sin(50 t) -0.25 t]$ (left picture). Mass  decay (right picture).}
\end{figure}

In Fig.~\ref{fig3} the modulated potential is
$V[x-10\sin(50 t) -0.25 t]$ (moving trap, with velocity $0.25$).
We plot the soliton profile and the soliton mass, which shows that radiative effects are very small.

\begin{figure}
\begin{center}
\begin{tabular}{c}
\includegraphics[width=6.4cm]{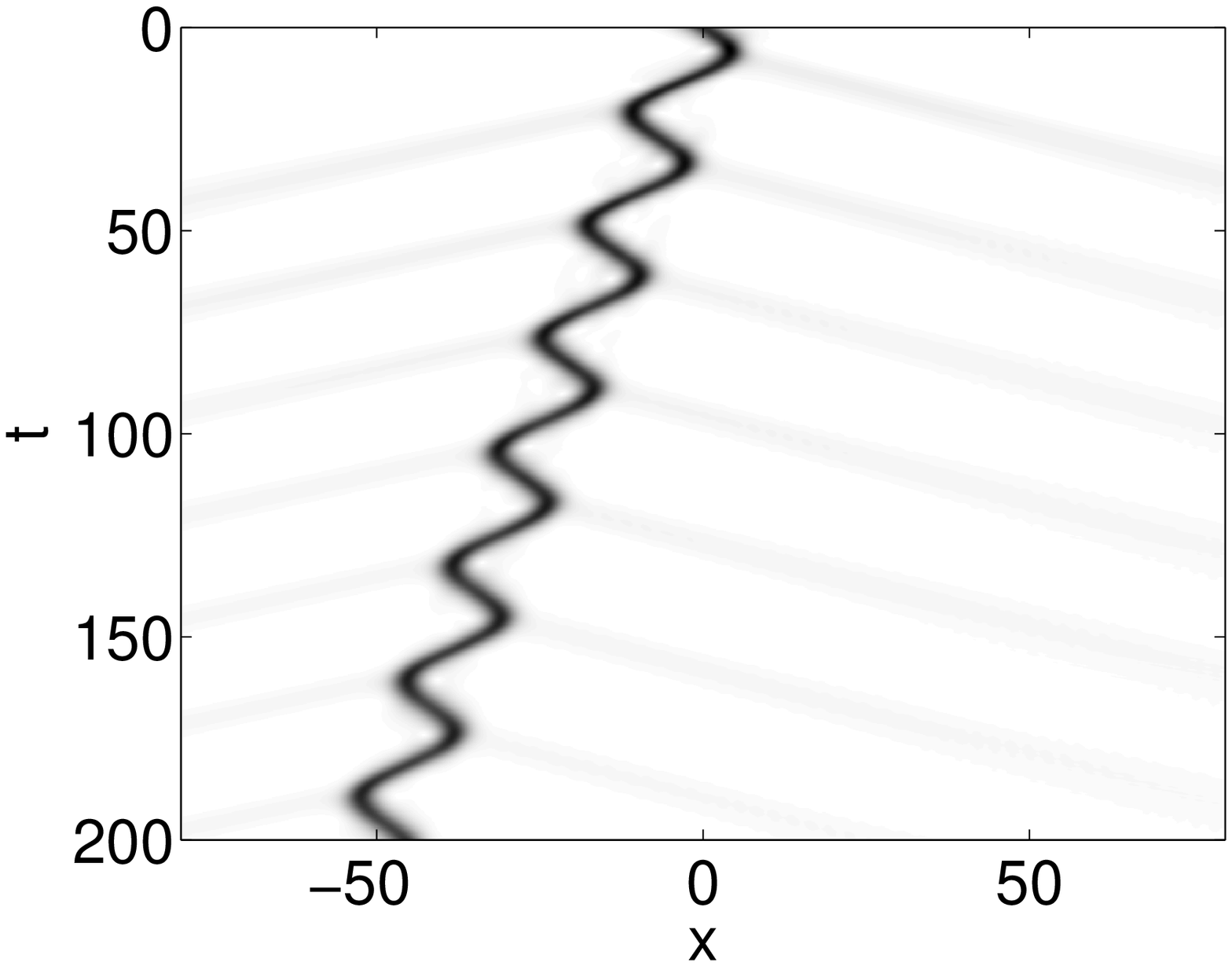}
\includegraphics[width=6.4cm]{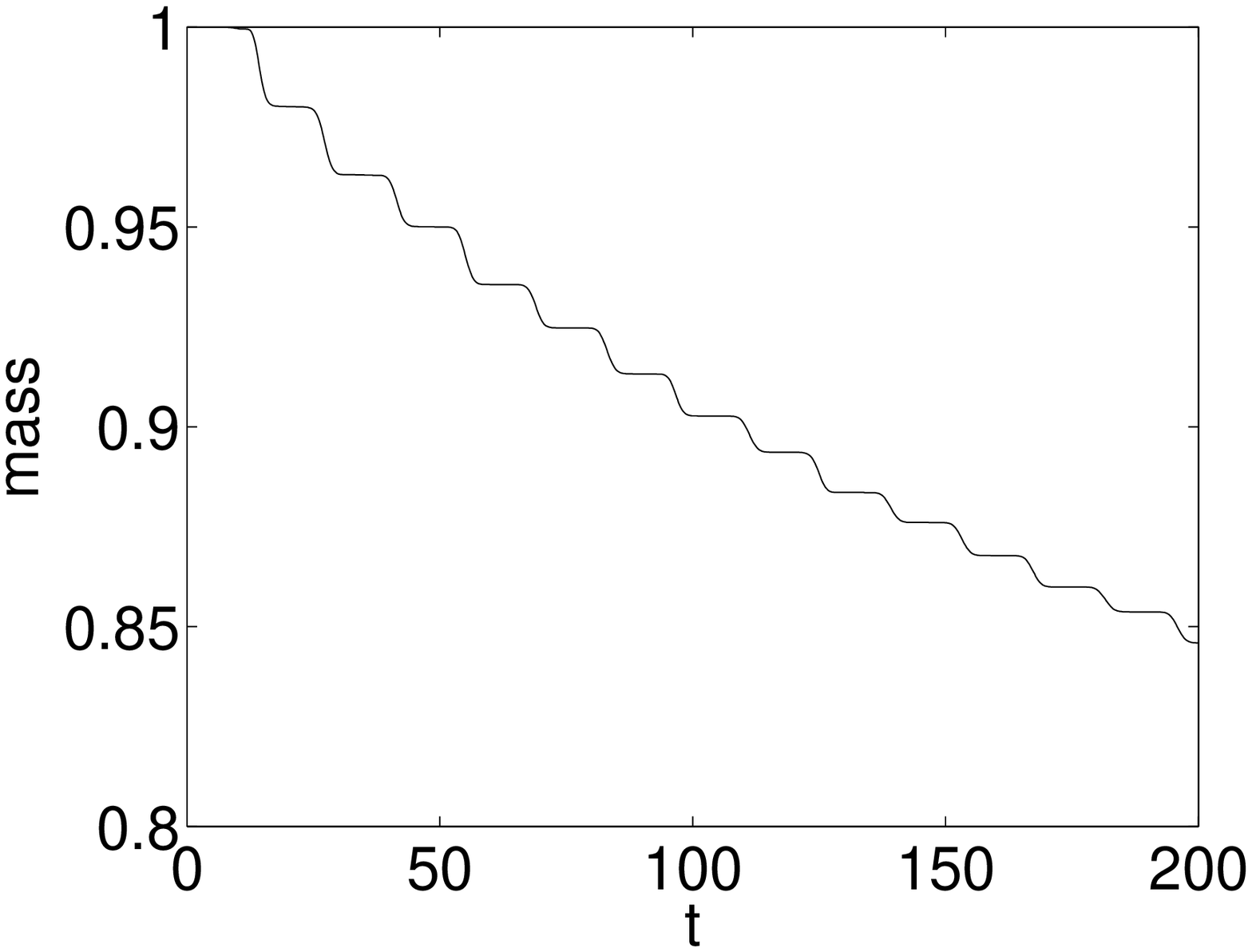}
\end{tabular}
\vspace*{-0.25in}
\end{center}
\caption{\label{fig4} Soliton profile in the moving  trap
$V[x - 10\sin(50 t)+0.25 t]$ (left picture). Mass  decay (right picture).}
\end{figure}

\begin{figure}
\begin{center}
\begin{tabular}{c}
\includegraphics[width=6.4cm]{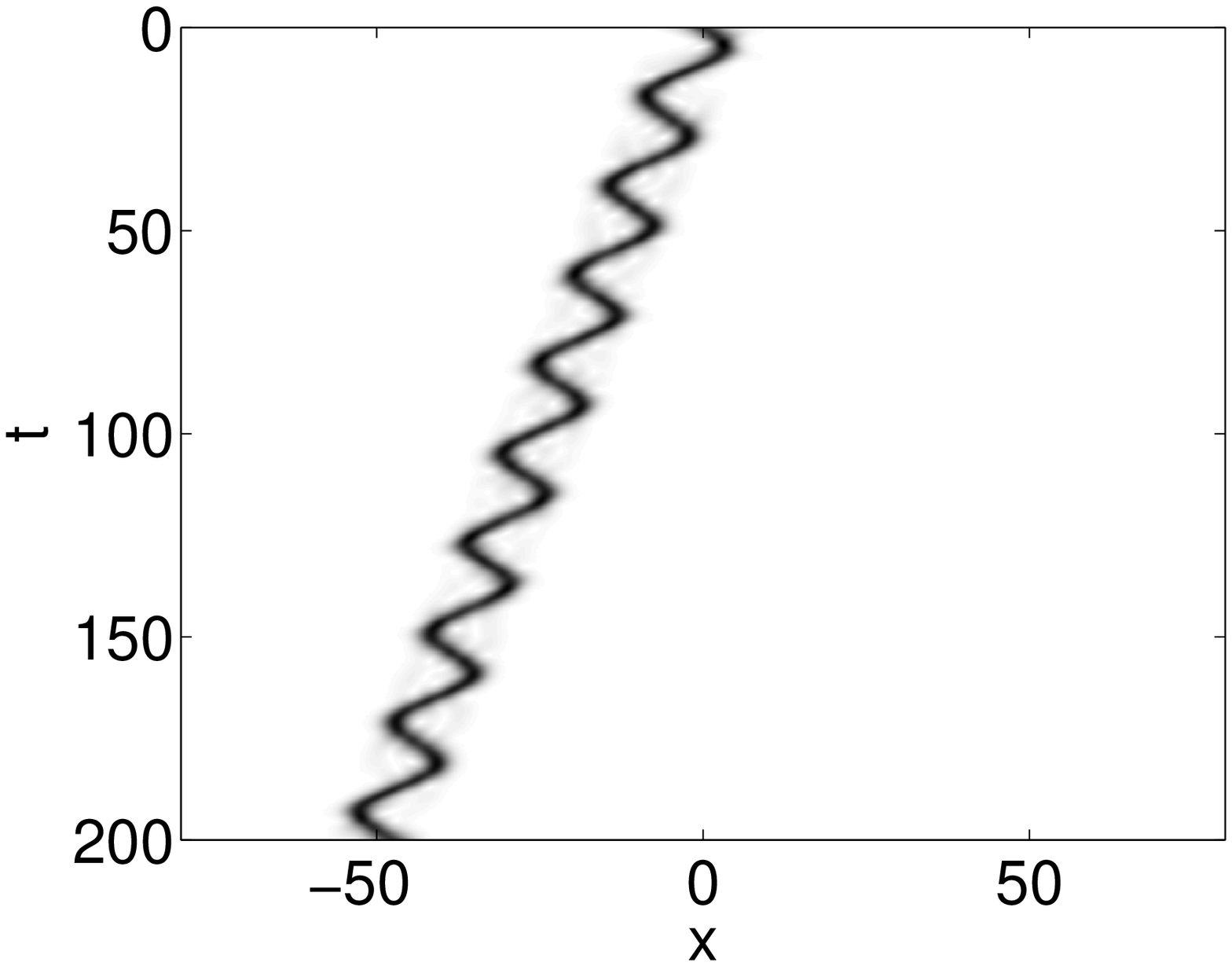}
\includegraphics[width=6.4cm]{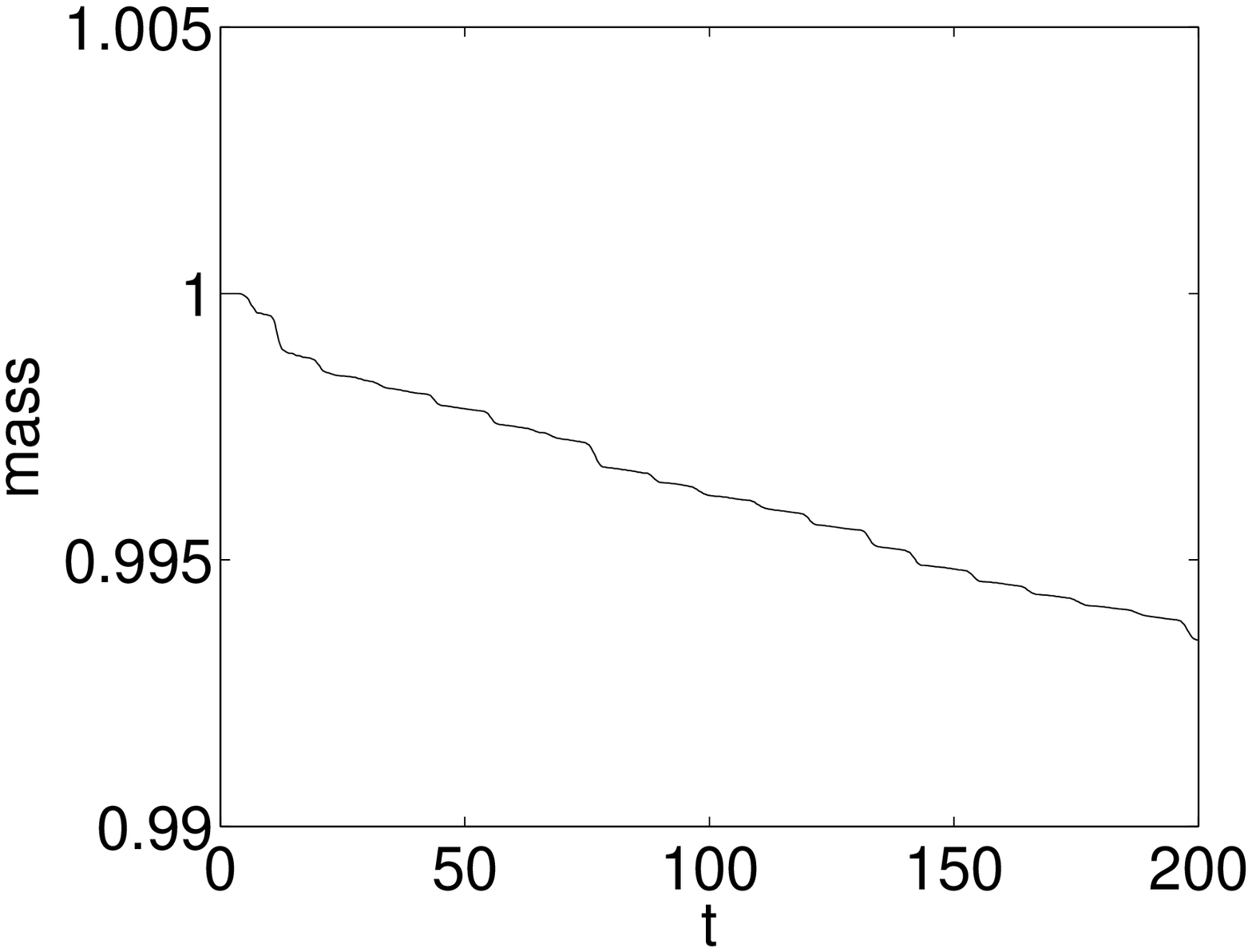}
\end{tabular}
\vspace*{-0.25in}
\end{center}
\caption{\label{fig5} Soliton profile in the moving  trap
$2V[x - 10\sin(50 t)+0.25 t]$ (left picture). Mass  decay (right picture).}
\end{figure}

In Fig.~\ref{fig4} the modulated potential is
$V[x-10\sin(50 t) +0.25 t]$ (moving trap, with velocity $-0.25$)).
Radiation effects are small, but not completely negligible.
It can be seen that some radiation is transmitted through the edges of the
effective double-barrier trap.
We then double the potential amplitude and consider
$2V[x-10\sin(50 t) +0.25 t]$.
The new simulation is shown in Fig.~\ref{fig5},
where radiation effects are seen to be negligible.

\begin{figure}
\begin{center}
\begin{tabular}{c}
\includegraphics[width=6.4cm]{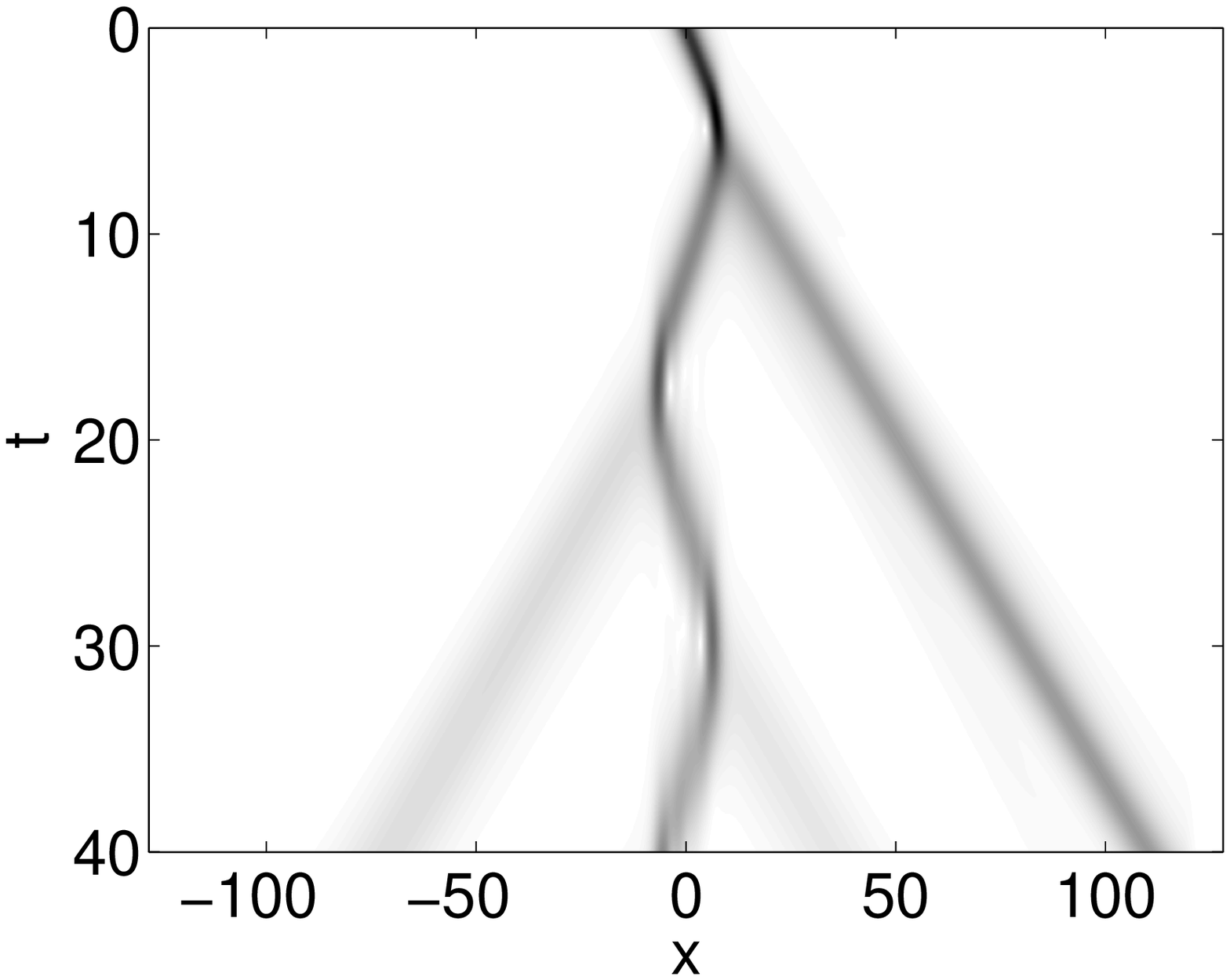}
\includegraphics[width=6.4cm]{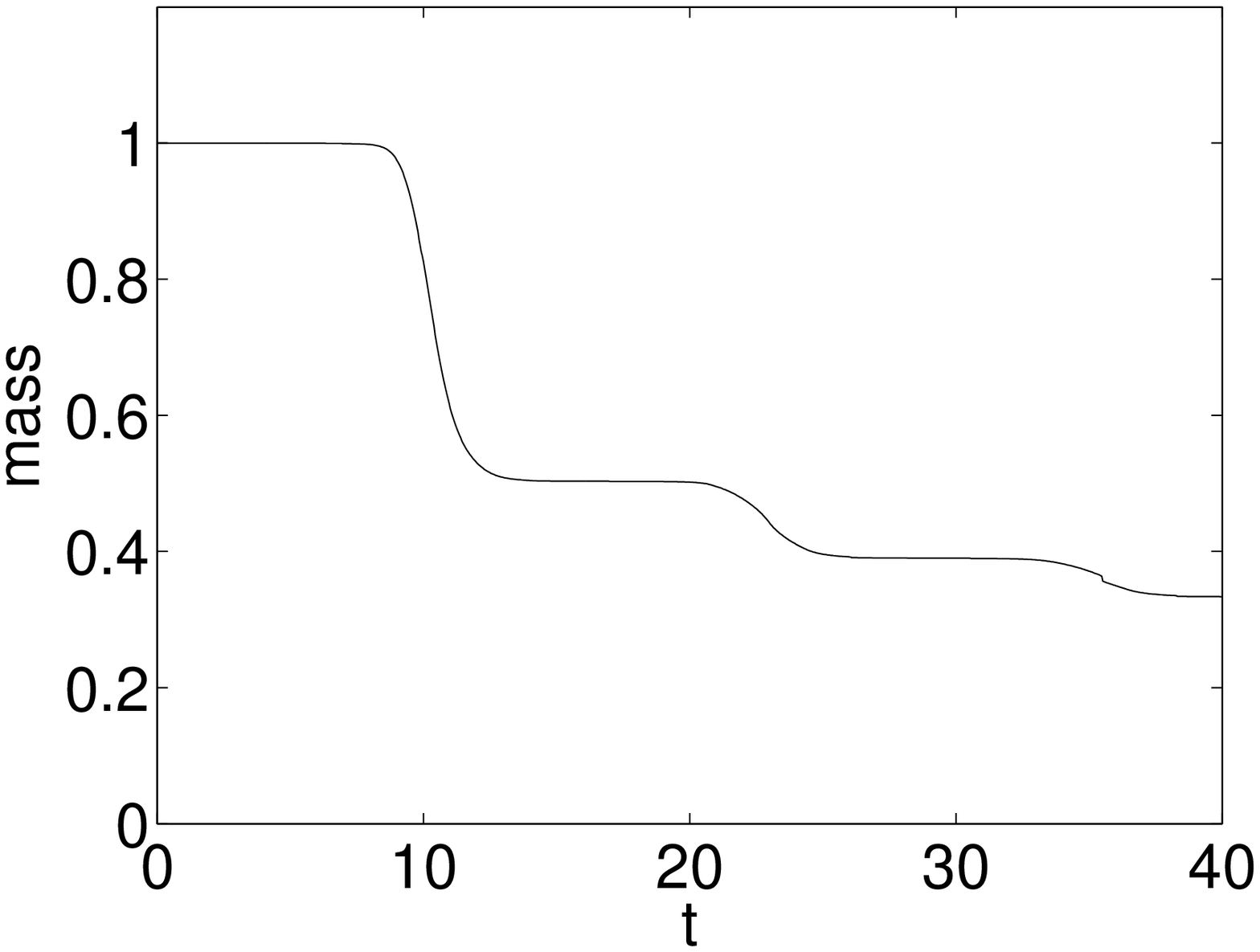}
\end{tabular}
\vspace*{-0.25in}
\end{center}
\caption{\label{fig6} Soliton profile in the stationary trap
$V[x - 10\sin(50 t)]$ (left picture). Mass  decay (right picture).
Here the initial velocity parameter is $\mu_0=0.5$.}
\end{figure}

In Fig.~\ref{fig6} we show the motion of a soliton with initial
parameters $\nu_0=0.25$ and $\mu_0=0.5$ (the soliton velocity is $2$).
Since we have $\Delta W_{\nu_0} \simeq 0.323$
and $\mu_c = 0.4$,  the quasi-particle approach
predicts unbounded motion.
The results of the numerical simulations confirm this prediction.
We can observe that half the initial soliton mass is transmitted through
the effective barrier during the first interaction, and the transmitted soliton has
a velocity that is larger than $4\mu_0$.

Let us estimate the typical values of the parameters for an experimental
configuration.
We consider
the case of a condensate with $^7$Li atoms in the a cigar-shaped trap
potential with $\omega_{\perp} = 2\pi \times 1$ kHz.
In this case, $\eps =2 \pi /50$ corresponds to
$1/T_p=50$ kHz. The barrier height is $V_{0} = 10\hbar\omega_{\perp}$. The
number of atoms can be taken as $N \sim 10^3$ for  the scattering
length $a_s = -0.2$~nm. The transverse oscillator length is
$a_{\perp} = 1.2~\mu$m, so the barrier width is $d \simeq 2~\mu$m and
the oscillations amplitude of the barrier position is $L \simeq 10~\mu$m. The velocity is normalized
by the sound velocity which is $c \simeq 4 $~mm/s, so the soliton
velocity parameter $\mu_0 =0.25$ corresponds to a velocity equal to $c$.

\section{Conclusion}
In conclusion we have studied the propagation of a bright matter wave
soliton through a barrier potential with rapidly oscillating position. We
have shown that the soliton can be dynamically trapped by such a potential.
It would be interesting to extend this model to the case of
oscillating barrier and standing well. In this case the effective
potential will have the form of two barriers separated by a well. Applying
the results obtained in Ref.~\cite{Azbel} we can expect for the
transmission of the nonlinear matter waves such phenomena as the
instability and quantum turbulence on the time scales of the
quantum dwell time in the well. Another physically relevant system
is the rapidly oscillating barrier in the well. The effective potential in this
case simulates a heterostructure and effective nonlinearities
can exhibit chaotic dynamics in the tunneling process \cite{Jona}.

In the case of a well with rapidly oscillating position the
effective potential will have a double-well form. In such a configuration
we can expect the existence of dynamically induced macroscopic
quantum tunnelling and localization phenomena, i.e. dynamically
generated bosonic Josephson junction. These problems require of
a separate investigation.

\end{document}